\documentclass[prd,aps]{revtex4}
\usepackage{amsmath}
\usepackage{amsfonts}
\usepackage{amssymb}

\newcommand{\eps}{\varepsilon}

\newcommand{\dpar}[4]{\frac{\partial #4}{\partial #1_{#2}^{#3}}}

\newcommand{\dmu}{\partial_{\mu}}
\newcommand{\dnu}{\partial_{\nu}}

\newcommand{\abs}[1]{|#1|}
\newcommand{\sint}{\mathcal{S}}

\newcommand{\fint}[2]{\int{\mathcal{D}#1\,#2}}

\begin{document}
\preprint{{hep-th/yymmnnn} \hfill {UCVFC-DF-17-2005}}
\title{The Topological Theory of the  Milnor Invariant
$\overline{\mu}(1,2,3)$}
\author{Lorenzo Leal$^{1}$
and Jes\'us Pineda $^{2}$}

\affiliation {1. Centro de F\'{\i}sica Te\'{o}rica y
Computacional, Facultad de Ciencias, Universidad Central de
Venezuela, AP 47270, Caracas 1041-A, Venezuela. \\
 2. Departamento
de F\'{\i}sica, Universidad Sim\'on Bol\'{\i}var,\\ Aptdo. 89000,
Caracas 1080-A, Venezuela.\\ }

\begin{abstract}
We study a topological Abelian gauge theory that generalizes the
Abelian Chern-Simons one, and that leads in a natural way to the
Milnor's link invariant $\overline{\mu}(1,2,3)$ when the classical
action on-shell is calculated.
\end{abstract}

\maketitle

\section{Introduction}
As it is well known, the vacuum expectation value (v.e.v.) of the
Wilson Loop (WL) (or of the product of several WLs) in the
Chern-Simons (CS) theory, produces knot (or link) invariants (LIs)
\cite{witten}. In the Abelian case, the invariants obtained are
the Self-Linking Number or the Gauss Linking Number (GLN),
depending on whether one deals with one or several Wilson lines.
In the non-Abelian case, on the other hand, the invariants
obtained are known to be knot or link polynomials, such as the
celebrated Jones polynomial \cite{witten}. Which polynomial
appears depends on the gauge group involved \cite{labastida}.

 The GLN of a pair of closed curves admits an analytical expression (see equation \ref{doce}) that has
a simple and appealing geometric interpretation: it represents the
oriented number of times that one of the curves flows through a
surface bounded by the other one. However, to obtain analytical
expressions in the non-Abelian case, one must resort to a
perturbative expansion of the v.e.v. of the WLs \cite{guadagnini},
since as far as it is known, polynomial invariants are not
expressible in analytical terms. This perturbative expansion
yields an infinite tower of analytical expressions for LIs of
increasing complexity. In general, it is not easy to elucidate
their geometric or topological meaning, however, the first three
of them have been explicitly calculated, and in some cases, a
geometrical interpretation is available
\cite{guadagnini,dibartolo,rozansky,mellor,lorenzo1,lorenzo3,labastida}.

 A question that naturally rises is
about the existence of an intermediate situation between the
Abelian and non-Abelian cases. Stated more precisely: is there any
topological field theory, other than the Abelian CS  (or the
Abelian BF theory) that yields \emph{exact} analytical expressions
for LIs, other than the GLN?.

 Beyond the theoretical interest that
this question could have, there is an increasing interest in the
description of phenomena that involve closed lines as relevant
structures (vortices and defects in condensed matter or fundamental physics,
loops in gauge theories and quantum gravity, polymer entanglements, among others examples).
Therefore, it could be useful to have at
one`s disposal new topological theories, that while going beyond
the Abelian Chern-Simons Theory and its associated GLN, do not
present the difficulties of the non-Abelian ones.

The purpose of this paper is to provide an example of such a
theory. As we shall see, the LI that the theory we are going to
consider produces is the Milnor's Linking Coefficient
$\overline{\mu}(1,2,3)$ , which is an invariant  associated to
links of at least three-components. This invariant follows the GLN
in an infinite family of link invariants discovered by Milnor
several decades ago \cite{milnor}.

The theory that we shall study can be seen as an effective Abelian
gauge and diffeomorphism invariant theory, that reproduces just
the second contribution of the perturbative expansion of a
non-Abelian topological one, namely, of the CS model coupled to
chromo-electrically charged particles (so called `Wong particles'
\cite{wong}). This Chern-Simons-Wong model has been recently
studied from a classical point of view \cite{lorenzo1, lorenzo3}.

The action that we are going to deal with comprises a pure
gauge-fields part, and terms representing the coupling of these
fields with external sources with support on closed curves. The
terms that correspond to the gauge-field's part coincide with
those of a recent article \cite{ferrari} that studies Chern-Simons
theories with
 non-semisimple group of symmetry. However, in
contrast with reference \cite{ferrari}, the interaction term that
we take is manifestly diffeomorphism invariant, and breaks the
non-Abelian gauge invariance of the former term down to an Abelian
gauge invariance (see equations (\ref{nueve}) and
(\ref{catorce})).

 In the discussion that follows, we shall adopt
the method of dealing with the classical (in the sense of non-
quantum-mechanical) theory to calculate LIs
\cite{lorenzo1,lorenzo2, lorenzo3}. Within this scheme, one solves
the equation of motion and calculates the on-shell (OS) action of
the topological theory coupled to external Wilson lines. The OS
action results to be a functional depending on the Wilson lines
that act as sources of the gauge theory and, since the theory is
metric independent, it is clearly a LI. For instance, when this
procedure is applied to the Abelian CS theory coupled to Wilson
lines, the OS action yields the GLN of the lines, just as in the
quantum case. This approach for obtaining LIs from classical field
theories can be rigorously proven and generalized to situations
where the symmetry group is other than the group of
diffeomorphisms  of the base manifold \cite{rafael, area}.
Although we shall focus mainly in this classical approach, we
shall also make some remarks about the "quantum method", which is
the procedure usually employed to study the relation between LIs
and topological theories.

\section{The action and the link invariant}\label{sec1}
The action that we shall study is given by
\begin{eqnarray}
\nonumber \sint&=&\int{d^{3}x\; \eps^{\mu\nu\rho}\big\{4\,A^{i}_{\mu}(x) \dnu a_{i\rho}(x)+ \frac{2}{3}\,\eps^{ijk} a_{i\mu}(x)a_{j\nu}(x)a_{k\rho}(x)\big\}} -2\,\int{d^{3}x\;T^{\mu x}_{i}A^{i}_{\mu}(x)}+\\
&&+\int{d^{3}x\;\int{d^{3}y\; \eps^{ijk}\,T^{\mu x,\nu y}_{i} a_{j\mu}(x)a_{k\nu}(y)}}. \label{uno}
\end{eqnarray}
Here, $ A^{i}_\mu (x)$ and $a^{i}_\mu (x)$ are two sets of
independent Abelian gauge fields, labelled by Latin letters
running from 1 to 3 (we use the summation convention of Einstein
also for these "internal" indexes). The first two terms would
correspond to the topological theory with non-semisimple gauge
group of symmetry introduced in reference \cite{ferrari}. The last
two terms in (\ref{uno}) involve the ``currents'' $T^{\mu
x}_{\gamma _{i}}$ and $T^{\mu x ,\nu y}_{\gamma _{i}}$, with
support on the three closed curves $\gamma _{i}$
\begin{equation}
T^{\mu y}_{i}= \oint_{\gamma_i}
dx^{\mu}\delta^{3}(x-y),\label{dos}
\end{equation}

\begin{equation}
T_{i}^{\mu x,\,\nu y} \equiv
\oint_{\gamma_i}dz^{\mu}\int_{0}^{z}dz'^{\nu}\delta^{3}(x-z)\delta^{3}(y-z').\label{tres}
\end{equation}

Under general coordinate transformations these objects behave as a
vector-density and a bi-local vector density respectively. They
obey the differential constraints
\begin{equation}
\partial_{\mu} T^{\mu y}_{\gamma_i}= 0,\label{cuatro}
\end{equation}
\begin{eqnarray}
\dpar{x}{}{\mu}{}T^{\mu x,\nu y}_{\gamma}&=& (-\delta^{3}(x-x_{0})+ \delta^{3}(x-y))T^{\nu y}_{\gamma}\\
\dpar{y}{}{\nu}{}T^{\mu x,\nu y}_{\gamma}&=& (\delta^{3}(y-x_{0})- \delta^{3}(y-x))T^{\mu x}_{\gamma}, \label{cinco}
\end{eqnarray}
and the algebraic constraint \cite{dibartolo}
\begin{equation}
T^{(\mu x,\nu y)}_{\gamma}\equiv \frac{1}{2}(T^{\mu x,\nu y}_{\gamma}+T^{\nu y,\mu x}_{\gamma})=T^{\mu x}_{\gamma} T^{\nu y}_{\gamma}. \label{seis}
\end{equation}
(observe that to the action (\ref{uno}) only contributes the
antisymmetric part (in $\mu x,\,\nu y$) of $T_{\gamma_a}^{\mu
x,\,\nu y}$).

 The ``loop coordinates''  $T^{\mu y}_{\gamma}$ and $T_{\gamma}^{\mu x,\,\nu
 y}$ are the first  members of an infinite sequence that
 arises when the path ordered exponential that defines the Wilson loop is
 expanded \cite{dibartolo}. As we shall see, the presence of the second
 ``loop-coordinate'' $T_{\gamma}^{\mu x,\,\nu y}$ is just what
 will lead us to obtain a LI beyond the GLN, which only depends on
 the first ``loop-coordinate'' $T^{\mu y}_{\gamma}$.

 Varying the action (\ref{uno}) with respect to $A_{\mu}^{i}$ and $a_{\mu}^{i}$ yields
\begin{equation}
\eps^{\mu\nu\rho}\dnu a_{i\rho}= \frac{1}{2}T^{\mu x}_{i}, \label{siete}
\end{equation}
\begin{equation}
\eps^{\mu\nu\rho} \dnu A_{\rho}^{i}(x)=- \frac{1}{2}\eps^{\mu\nu\rho}\eps^{ijk} a_{j\nu}(x)a_{k\rho}(x)+\frac{1}{2}\int{d^{3}y\; \eps^{ijk}T^{[\mu x,\nu y]}_{j}a_{k\nu}(y)}. \label{ocho}
\end{equation}
These equations are just the $0-th$ and first order contributions
to the $SU(2)$ Chern-Simons-Wong equations of motion that were
studied in references \cite{lorenzo1, lorenzo3}. In that approach,
the fields $A_{\mu}^{i}$ and $a_{\mu}^{i}$ correspond,
respectively, to the first and $0-th$ contributions of a
perturbative expansion for the non-Abelian potential
\cite{lorenzo1,lorenzo3}.

Since $T^{\mu y}_{\gamma_i}$ is divergenceless, equation
(\ref{siete}) is consistent. This reflects the invariance of the
action under the gauge transformations
\begin{equation}
A_{\mu}^{i} \longrightarrow A_{\mu}^{i} + \partial_{\mu}
\Lambda^{i}. \label{nueve}
\end{equation}
The consistency of equation (\ref{ocho}) is more involved. Taking
the divergence of both sides of this equation yields
\begin{equation}
0=2\eps^{\mu\nu\rho}\eps^{ijk} \dmu (a_{j\nu}(x) a_{k\rho}(x))-
\eps^{ijk}\int{d^{3}y\,a_{k\nu}(y)\dpar{x}{}{\mu}{}T^{[\mu x,\nu
y]}_{j} }. \label{diez}
\end{equation}
Using the differential constraints (\ref{cinco}) and the equation
of motion (\ref{siete}), we obtain
\begin{equation}
\nonumber\eps^{ijk}\delta^{3}(x-x_{j}(0)) \oint_{j}{dx^{\nu}\,\oint_{k}{dz^{\beta}\,
 \eps_{\nu\alpha\beta}\,\frac{(x-z)^{\alpha}}{\abs{x-z}^{3}}}}=\eps^{ijk}\delta^{3}(x-x_{j}(0))L(j,k)=0, \label{once}
\end{equation}
were
\begin{equation}
L(i,j)=\frac{1}{4\pi}\, \oint_{\gamma_{i}}{dz_{i}^{\beta}\;\oint_{\gamma_{j}}{dy^{\mu}_{j}\; \eps_{\mu\beta\gamma}\frac{(z-y)^{\gamma}}{\abs{z-y}^{3}}}},\label{doce}
\end{equation}
is the GLN between the curves \(\gamma_{i}\) and \(\gamma_{j}\). In the case where these
curves do not intersect each other, equation (\ref{once}) demands
that
\begin{equation}
L(i,j) = 0 \quad \forall \,\, i,j.
\label{trece}
\end{equation}
From this result we obtain that the theory is consistent whenever
the curves  are not linked in the sense of the GLN. This does not
mean that the curves are equivalent to the trivial link (the
unlink). For instance, the Borromean Rings are a well known set of
three curves whose GLNs vanish, although they are indeed entangled
\cite{rolfsen}. There are more complex entanglement patterns
associated with CS theory than those measured by the GLN.

The consistency condition (\ref{trece}) is also related to a gauge
symmetry of the theory. A direct calculation shows that the action
(\ref{uno}) is invariant under the transformations

\begin{equation}
a_{i\mu}\rightarrow a_{i\mu}+\dmu\Omega_{i}, \label{catorce}
\end{equation}
provided that the consistency condition (\ref{trece}) is
fulfilled. Thus, we see that both sets of fields  $A_{i}$ and
$a_{i}$ must be Abelian gauge fields for the theory to be
consistent.

On the other hand, there is no need of introducing a metric in the
manifold to construct the action, as can be easily verified.
Hence, the theory is metric independent. Since it is also
generally covariant, it is a topological theory, just like its
cousins the Abelian and non-Abelian Chern-Simons theories. Hence,
following references \cite{lorenzo1, lorenzo2, lorenzo3}, we
conclude that the on-shell action $S_{os}$ of the theory should
only depend on topological features of the curves appearing in the
action, i.e., it should be a link invariant. Let us see how this
happens.

The solution of equation (\ref{siete}) is given by
\begin{equation}
a_{i\mu}(x)=-\Big(\frac{1}{2}\Big) \frac{1}{4\pi}\oint_{\gamma_{i}}{dz^{\rho}\; \eps_{\mu\nu\rho}\frac{(x-z)^{\nu}}{\abs{x-z}^{3}}}. \label{quince}
\end{equation}
Equation (\ref{ocho}) can also be integrated as easily as the
former one, but in order to calculate $S_{os}$ it suffices to
substitute the left hand side of (\ref{ocho}) and expression
(\ref{quince}) into (\ref{uno}). The result is then

\begin{eqnarray}
\nonumber S^{(1)}(1,2,3)&=&-\frac{1}{2}\int{d^{3}x\; \eps^{\mu\nu\rho}a_{1\mu}(x)a_{2\nu}(x)a_{3\rho}(x)}-
\frac{1}{2}\int{d^{3}x\;\int{d^{3}y\;\Bigl(T^{[\mu x,\nu y]}_{1} a_{2\mu}(x)a_{3\nu}(y)}+}\\
&&+T^{[\mu x,\nu y]}_{2} a_{3\mu}(x)a_{1\nu}(y)+T^{[\mu x,\nu y]}_{3} a_{1\mu}(x)a_{2\nu}(y)\Bigr). \label{dieciseis}
\end{eqnarray}

Equation (\ref{dieciseis}) corresponds to an analytical expression
for the Milnor's Linking Coefficient $\overline{\mu}(1,2,3)$
\cite{milnor}. In \cite{lorenzo3} it was shown by explicit
calculation that this expression, despite its appearance, is
metric independent, as it should be. An interpretation of its
geometrical meaning can be found in references \cite{mellor,
lorenzo3}.

A sketch of the interpretation of this result would be as follows:
the first term in equation (\ref{dieciseis}) measures how many
times  three arbitrary surfaces whose boundaries are the three
curves of the theory (known as Seifert surfaces \cite{rolfsen})
intersect at a common point. The second term counts the oriented
number of times that one of the curves  crosses first the surface
bounded by the second curve and then the surface bounded by the
last one. The fact that expression (\ref{dieciseis}) keeps memory
of the order in which each curve flows through the surfaces
attached to the other curves is what distinguishes this invariant
from the GLN. This feature makes \(\overline{\mu}(1,2,3)\) a
natural ``next level'' of complexity LI when compared to the GLN.
Obviously, further developments along  these lines  might provide
even more (and more interesting) link invariants.

To conclude let us briefly discuss the quantum formulation of the
theory, within the Feynman path-integral framework.  We consider
the functional integral

\begin{equation}
W(\gamma_{i})= \fint{A}{\fint{a}{\exp{(-S)}}}. \label{diecisiete}
\end{equation}

It should be noticed that action (\ref{uno}) already depends
on the "Wilson lines" $\gamma_{i}$. This dependence is mandatory
to preserve the Abelian gauge-invariance given by (\ref{nueve})
and (\ref{catorce}). This contrasts with what occurs in the usual
Abelian Chern-Simons theory and the non-Abelian one, where gauge
invariance does not demand the coupling with external Wilson
lines. Integrating out the fields $A$ produces a functional "delta
function"
\begin{equation}
\fint{A}\exp{\Big\{\int{d^{3}x\;A^{i}_{\mu}(x)\big\{
4\,\eps^{\mu\nu\rho}\dnu a_{i\rho}(x)-2T^{\mu
x}_{i}\big\}}\Big\}}\propto \;\delta[4\,\eps^{\mu\nu\rho}\dnu
a_{i\rho}(x)-2T^{\mu x}_{i}], \label{dieciocho}
\end{equation}
that when substituted into (\ref{diecisiete}) enforces the $a$
fields to take their on-shell values given by (\ref{quince}).
Hence, the result is
\begin{equation}
W(\gamma_{i})= C \exp (S_{os}), \label{diecinueve}
\end{equation}
where $C$ is a constant and $S_{os}$ is the link invariant given
in equation (\ref{dieciseis}).

In the preceding discussion we have ignored that, indeed, gauge
invariance leads to infinities in the Feynman path-integrals that
should be properly handled. This can be done, for instance, by
employing the Faddeev-Popov method \cite{faddeev} in the usual way.
The result still is given by (\ref{diecinueve}), as can be readily
checked.

Summarizing, we have presented a topological model that
``interpolates'' between Abelian and Non-Abelian Chern-Simons
theory, in the sense that it leads to a link invariant that goes
beyond the GLN yielded by the Abelian theory, but otherwise
retaining the property of being an exactly soluble model, unlike
the non-Abelian one. The link invariant so obtained corresponds to
the Milnor linking coefficient $\overline{\mu}(1,2,3)$.

This work was supported by Proyect PI 03-00-6316-2006 of \emph
{Consejo de Desarrollo Cient\'{\i}fico y Human\'{\i}stico},
Universidad Central de Venezuela, Caracas, VENEZUELA, and by
Project G-2001000712 of FONACIT, VENEZUELA.

\end{document}